\documentclass[10pt]{iopart}
\usepackage{iopams}
\usepackage{hyperref}
\hypersetup{
colorlinks=true,
linkcolor=blue,
citecolor=blue,
urlcolor=blue}
\newcommand{\intprod}{\mbox{$\;
\put(0,0){\line(1,0){.9}}\put(.9,0){\line(0,1){1.6}} \; \, \, $}}

\begin{document}
 
\title[Symmetries of first-order Lovelock gravity]{Symmetries of
first-order Lovelock gravity} \author{Merced Montesinos$^1$, Rodrigo
Romero$^1$ and Bogar D\'iaz$^2$} \address{$^1$ Departamento de
F\'isica, Cinvestav, Avenida Instituto Polit\'ecnico Nacional 2508,
San Pedro Zacatenco, 07360, Gustavo A. Madero, Ciudad de M\'exico,
Mexico} \address{$^2$ Departamento de Matem\'aticas, Instituto de
Ciencias, Benem\'erita Universidad Aut\'onoma de Puebla, Ciudad
Universitaria, 72570, Puebla, Puebla, Mexico}
\eads{\mailto{merced@fis.cinvestav.mx},
  \mailto{rromero@fis.cinvestav.mx} and  
\mailto{bdiaz@fcfm.buap.mx}}

\begin{abstract} 
We apply the converse of Noether's second theorem to the first-order $n$-dimensional Lovelock action, considering the frame rotation group as both $SO\left(1,n-1\right)$ or as $SO(n)$. As a result, we get the well-known invariance under local Lorentz transformations or $SO(n)$ transformations and diffeomorphisms, for odd- and even-dimensional manifolds. We also obtain the so-called `local translations' with nonvanishing constant $\Lambda$ for odd-dimensional manifolds when a certain relation among the coefficients of the various terms of the first-order Lovelock Lagrangian is satisfied. When this relation is fulfilled, we report the existence of a new gauge symmetry emerging from a Noether identity. In this case the fundamental set of gauge symmetries of the Lovelock action is composed by the new symmetry, local translations with $\Lambda \neq 0$ and local Lorentz transformations or $SO(n)$ transformations. The commutator algebra of this set closes with structure functions. We also get the invariance under local translations with $\Lambda=0$ of the highest term of the Lovelock action in odd-dimensional manifolds. Furthermore, we report a new gauge symmetry for the highest term of the first-order Lovelock action for odd-dimensional manifolds. In this last case, the fundamental set of gauge symmetries can be considered as Poincar\'e or Euclidean transformations together with the new symmetry. The commutator algebra of this set also closes with structure functions.
\end{abstract}

\maketitle

\section{Introduction}\label{sec:intro}

Noether's theorems and their converses have become the fundamental mathematical tools for linking symmetries with conservation laws for dynamical systems whose equations of motion can be obtained from a variational action principle \cite{Noether1918, Noether1918Eng} (see also \cite{Bessel-Hagen1921}). In particular, the converse of Noether's second theorem can be used to reveal a gauge symmetry of a theory from a Noether identity and, even when the symmetries are already known, the theorem might be used to reformulate them by constructing a different Noether identity. For instance, in \cite{Montesinos2017} the converse of Noether's second theorem was applied to the $n$-dimensional Palatini and Holst actions, both with cosmological constant, showing that in each case a new gauge symmetry that is the generalization to higher dimensions ($n>3$) of the so-called `local translations' \cite{Witten:1988hc,ACHUCARRO198689} (see also \cite{carlip2003quantum}) for the three-dimensional Palatini action with cosmological term naturally emerges. The new symmetry and local Lorentz transformations can be taken as the fundamental gauge symmetries of first-order general relativity. The algebra of these symmetries turns out to be \textit{open} \cite{Henneaux1990,Henneaux1994}. As a consequence, an infinitesimal diffeomorphism can be expressed as a linear combination of the new gauge symmetry and local Lorentz transformations with field-dependent gauge parameters, up to terms involving the variational derivatives, and therefore diffeomorphisms become a derived symmetry. The new gauge symmetry has also been extended to include the coupling of Yang-Mills and fermion fields to the Palatini and Holst actions with cosmological constant in \cite{Montesinos2018}. 

All of this suggests that it is worth to explore generalizations of general relativity from the perspective of Noether's second theorem. In this paper we focus on Lovelock theory, which is defined by an action over an $n$-dimensional manifold, whose variational derivatives may be considered as generalizations of the Einstein tensor \cite{Lovelock1971}. In the metric formalism, the equations of motion of Lovelock theory are second order at most and this fact avoids causality problems at the classical level. Lovelock gravity can also be analyzed in the first-order formalism \cite{Zanelli2012}. In this framework it is pretty obvious that the space of solutions of Lovelock gravity allows configurations with nonvanishing 
torsion \cite{BlackHoles1,Troncoso2000,Black5D}. Nevertheless, the torsion-free sector of the space of solutions propagates the same number of physical degrees of freedom as general relativity \cite{TorsionDegrees} and this sector has been exhaustively studied in the literature (see, for instance, \cite{ReviewSolutions} and references therein). Therefore, Lovelock gravity provides a non-trivial toy model to confront its space of solutions with the one of general relativity and get deep insights about the nature of gravity. In this paper, we will study off-shell Lovelock theory and, in particular, we do not restrict the analysis to torsion-free configurations.

Our work is organized as follows: in section \ref{sec:lovelockaction} we introduce the notation and conventions that we will use in this paper. We also review the main aspects of Lovelock gravity in the first-order formalism, and present the framework in which we will apply the converse of Noether's second theorem. Then, in section \ref{sec:Gauge} we derive the well-known gauge symmetries of the Lovelock action by means of the converse of Noether's second theorem. Lovelock action is by construction invariant under local Lorentz transformations (or $SO(n)$ transformations depending on the frame rotation group) and diffeomorphisms, and so it is not surprising that we get the Noether identities related to these symmetries. However, it is interesting to note that the symmetry that arises from our analysis is a linear combination of diffeomorphisms and local Lorentz transformations ($SO(n)$ transformations) rather than simply diffeomorphisms. Furthermore, in three-dimensions Lovelock action coincides with three-dimensional first-order general relativity and, as we already mentioned, it is very well-known that this theory has a gauge symmetry referred to as `local translations' \cite{Witten:1988hc, ACHUCARRO198689} (see also \cite{carlip2003quantum}). In this section we also explore the existence of a generalization of this symmetry to the $n$-dimensional first-order Lovelock action. We obtain that local translations with $\Lambda \neq 0$\footnote{In this context, $\Lambda$ is not the cosmological constant, but a proportionality constant whose definition is given below, in equation \eref{eq:conditionrecurrence}.} are a symmetry of the Lovelock action only in odd dimensions when a certain condition between the many terms of the Lovelock action is satisfied. At the end of this section we show that in odd dimensions the highest term of the Lovelock action is quasi-invariant under local translations with $\Lambda = 0$. In section \ref{subsec:NewsymmetryLnot0} we show that for odd dimensions and when the previous condition for the coefficients of the Lovelock action is fulfilled, we can construct a Noether identity from which we get a new gauge symmetry. In this case, the fundamental set of gauge symmetries of the Lovelock action is composed by the new symmetry together with the local translations with $\Lambda \neq 0$ and local Lorentz transformations (or $SO(n)$ transformations). In order to get information about the new symmetry, we compute the commutator algebra of this set, obtaining a closed algebra which closes with structure functions. For the sake of completeness, we also compute the commutator algebra of two equivalent complete sets of gauge symmetries for this case. After that, we show that diffeomorphisms become a derived symmetry, since any infinitesimal diffeomorphism can be written as a linear combination of the new symmetry, local Lorentz transformations (or $SO(n)$ transformations) and Local translations with $\Lambda \neq 0$ and field-dependent gauge parameters. Finally, in section \ref{sec:newsymmetry} we report the existence of a new internal gauge symmetry for the highest term in the Lovelock action for odd-dimensional manifolds. This symmetry results from the construction of a Noether identity, which according to the converse of Noether's second theorem, encodes a gauge symmetry of the theory. The new symmetry and the Poincar\'e (or Euclidean ones) can be taken as the fundamental set of gauge symmetries of this particular action. From this perspective, diffeomorphisms also become a derived symmetry because any infinitesimal diffeomorphism can be written in terms of these symmetries with field-dependent gauge parameters. Also, in this section we report the commutator algebra of the symmetries of the highest Lovelock term, which closes with structure functions.

\section{Lovelock action in the first-order formalism} \label{sec:lovelockaction}
Let us establish the notation and conventions used throughout this paper. We consider an $n$-dimensional orientable manifold $\mathcal{M}^{n}$ with $n \geqslant 3$. The frame rotation group is denoted by $SO(\sigma)$ where $SO(-1) \equiv SO\left(1,n-1\right)$ for Lorentzian manifolds ($\sigma=-1$) and $SO(+1) \equiv SO(n)$ for Euclidean ones ($\sigma=+1$) \textit{i.e.}, $e^{I}$ is an orthonormal frame of $1$-forms. For the rest of this work, when we talk about the `Lorentz group' we are referring to both cases. The indices $I_{1},\ldots, I_{n}$ run from $1$ to $n$ and are raised and lowered with the metric $(\eta_{IJ}) = \mathrm{diag}\left(\sigma,1,...,1\right) $, and the volume form is given by $\eta=(1/n!)\epsilon_{I_{1} \cdots I_{n}} e^{I_{1}} \wedge \cdots \wedge e^{I_{n}}$, where the rotation group tensor $\epsilon_{I_{1} \cdots I_{n}}$ is totally antisymmetric and satisfies $\epsilon_{ 1 \cdots n}=1$.

Lovelock gravity \cite{Lovelock1971} emerges from looking for an action principle that leads to a generalization of the Einstein tensor in dimensions higher than three. The Lovelock action is given by \cite{Lovelock1971}(see also \cite{Zanelli2012})
\begin{equation}
  \label{eq:lovelockaction}
  S_{n}[e,\omega] = \int \sum_{p=0}^{\left[n/2\right]} a_{p} L_{n}^{p},
\end{equation}
which is defined over $\mathcal{M}^{n}$. Here, $a_{p}$ are arbitrary constants, $[c]$ denotes the integer part of the real constant $c$, and $L_{n}^{p}$ is the $n$-form
\begin{equation}
  \label{eq:lovelockterms}
  \fl L_{n}^{p}= \kappa \epsilon_{I_{1} I_{2} \cdots I_{2p-1} I_{2p} I_{2p+1} \cdots I_{n}} R^{I_{1}I_{2}} \wedge \cdots \wedge R^{I_{2p-1} I_{2p}} \wedge e^{I_{2p+1}} \wedge \cdots \wedge e^{I_{n}},
\end{equation}
where $\kappa$ is a constant whose dimensions depend on $n$, and $R^{IJ}$ is the curvature of the $SO(\sigma)$ connection $\omega^{IJ}(=-\omega^{JI})$, defined as $R^{IJ}:=d\omega^{IJ}+\omega^{I}\,_{K} \wedge \omega^{KJ}$. Notice that $p$ counts the number of curvature factors in $L_{n}^{p}$.

By computing the variation of the Lovelock action \eref{eq:lovelockaction} with respect to the independent fields $e^{I}$ and $\omega^{IJ}$, we get
\begin{equation}
  \label{eq:definitionvariationalderivatives}
  \delta S_{n} = \int \left[ \mathcal{E}_{I}\wedge \delta e^{I} + \mathcal{E}_{IJ}\wedge \delta \omega^{IJ} + d\left ( \theta_{IJ} \wedge \delta \omega^{IJ} \right ) \right],
\end{equation}
from which we can read off the variational derivatives, and the $(n-2)$-form $\theta_{IJ}$ involved in the surface term, namely 
\begin{eqnarray}
  \label{eq:eulere}
  \mathcal{E}_{I}=  \sum_{p=0}^{\left[\left(n-1\right)/2\right]}
  {a_{p}\mathcal{E}_{I}^{p}},
\\  \label{eq:eulerw}
  \mathcal{E}_{IJ}=  \sum_{p=1}^{\left[\left(n-1\right)/2\right]}
  {a_{p}\mathcal{E}^{p}_{IJ}},
\\
  \label{eq:theta}
  \theta_{IJ} = \sum_{p=1}^{\left[n/2\right]}
  {a_{p}\theta^{p}_{IJ}},
\end{eqnarray}
where $\mathcal{E}^{p}_{I}$ and $\mathcal{E}^{p}_{IJ}$ denote the variational derivatives of each term $\int L_{n}^{p}$ in the sum \eref{eq:lovelockaction}, and $\theta_{IJ}^{p}$ is a $(n-2)$-form given by
\begin{eqnarray}
   \mathcal{E}_{I}^{p} =
  & \left(-1\right)^{n-1} \kappa\left(n-2p\right)
    \epsilon_{I I_{2} I_{3} \cdots I_{2p}I_{2p+1}I_{2p+2} \cdots  I_{n}} 
    \nonumber\\
  & \times R^{I_{2}I_{3}} \wedge \cdots \wedge R^{I_{2p}I_{2p+1}} \wedge
    e^{I_{2p+2}} \wedge \cdots \wedge e^{I_{n}}, 
    \label{eq:eulerep}
  \\
  \mathcal{E}_{IJ}^{p} =
  & \left(-1\right)^{n-1} \kappa p \left(n-2p\right)
    \epsilon_{I J I_{3} I_{4}\cdots I_{2p-1} I_{2p} I_{2p+1} I_{2p+2}\cdots I_{n}}
    \nonumber\\
  & \times R^{I_{3}I_{4}} \wedge \cdots \wedge R^{I_{2p-1}I_{2p}} \wedge
    De^{I_{2p+1}} \wedge
    e^{I_{2p+2}}\wedge \cdots \wedge e^{I_{n}},
    \label{eq:eulerwp}
  \\
  \theta_{IJ}^{p} =
  & \left(-1\right)^{n} \kappa p
    \epsilon_{IJ I_{3}I_{4}\cdots I_{2p-1}I_{2p}I_{2p+1}\cdots I_{n}}
    \nonumber\\
  & \times R^{I_{3}I_{4}} \wedge \cdots \wedge R^{I_{2p-1}I_{2p}} \wedge
    e^{I_{2p+1}} \wedge \cdots \wedge e^{I_{n}}.
    \label{eq:thetap}
\end{eqnarray}
Here $D$ is the $SO(\sigma)$ covariant derivative computed with $\omega^{IJ}$. The upper and lower limits in the sums \eref{eq:eulere}-\eref{eq:theta} come from the following facts:

(i) The first term ($p=0$) of the Lovelock lagrangian is
\begin{equation}
  \label{eq:Firsttermlagrangian}
  L_{n}^{0} = \kappa
  \epsilon_{I_{1} \cdots I_{n}}e^{I_{1}}\wedge \cdots \wedge e^{I_{n}},
\end{equation}
and so, we observe that this term does not have a variational derivative with respect to the connection $\omega^{IJ}$.

(ii) On the other hand, for even $n$ the last term of the Lovelock lagrangian is proportional to the Euler form
\begin{equation}
  \label{eq:EulerForm}
  L_{n}^{n/2} = \kappa \epsilon_{I_{1}I_{2} \cdots I_{n-1}I_{n}}
  R^{I_{1} I_{2}} \wedge \cdots \wedge R^{I_{n-1} I_{n}}.
\end{equation}
Hence, we note that $L_{n}^{n/2}$ does not have a variational derivative with  respect to $e^{I}$. The variation of the action constructed with \eref{eq:EulerForm} with respect to the connection is given by
\begin{equation}
  \label{eq:intEulerForm}
  \delta \int L_{n}^{n/2} =
  \kappa \int{ \left( \mathcal{E}^{n/2}_{IJ}
    \wedge \delta \omega^{IJ} +
    d\left(\theta_{IJ}^{n/2} \wedge \delta\omega^{IJ} \right)\right)},
\end{equation}  
with
\begin{eqnarray}
  \label{eq:eulerwmaximal}
  \mathcal{E}_{IJ}^{n/2}=
  \left(-1\right)^{n-1} \kappa (n/2)
  D \left( \epsilon_{IJI_{3}I_{4}\cdots I_{n-1}I_{n}}
  R^{I_{3}I_{4}} \wedge \cdots \wedge R^{I_{n-1}I_{n}}\right),
  \\\label{eq:thetamaximal}
  \theta_{IJ}^{n/2} =
  \left(-1\right)^{n-1} \kappa
  (n/2)\epsilon_{IJI_{3}I_{4}\cdots I_{n-1} I_{n}}
  R^{I_{3}I_{4}} \wedge \cdots \wedge R^{I_{n-1}I_{n}}.
\end{eqnarray}
We observe from \eref{eq:eulerwmaximal} that $\mathcal{E}_{IJ}^{n/2}=0$ because of the Bianchi's identity $DR^{IJ}=0$.

Therefore, the variational derivatives of $ \int L^{n/2}_{n}$ vanish, which means that this term is quasi-invariant under any infinitesimal transformation of the independent fields $e^{I}$ and $\omega^{IJ}$. This is so because the term \eref{eq:EulerForm} is topological.

\textit{Local Lorentz transformations and diffeomorphisms}. Each term $\int L_{n}^{p}$ is clearly Lorentz invariant (since $L_{n}^{p}$ is the wedge product of Lorentz tensors) and, under an active diffeomorphism $\Psi:\mathcal{M}^{n}\rightarrow\mathcal{M}^{n}$, $L_{n}^{p}$ satisfies $\Psi^{*}L_{n}^{p}\left(e,\omega\right)=L_{n}^{p}\left(\Psi^{*}e,\Psi^{*}\omega\right)$. Therefore, by Stokes theorem, each term $\int L_{n}^{p}$ is diffeomorphism-invariant as well. We can restate these properties in the framework of infinitesimal transformations, which are naturally adapted to the converse of Noether's second theorem.

An infinitesimal transformation of the fields depending on arbitrary functions and their derivatives is said to be a gauge symmetry of the action if the action remains invariant up to a total derivative under the transformation of the fields \cite{Noether1918,Noether1918Eng,Bessel-Hagen1921}. The infinitesimal local Lorentz and diffeomorphisms transformations are
\begin{eqnarray}
  \label{eq:varLorentzgroup}
  \mathrm{Lorentz:} \quad \delta_{\tau} e^{I} = \tau^{I}\,_{J}e^{J},
  \quad \delta_{\tau}\omega^{IJ}=-D\tau^{IJ},
  \\
  \label{eq:vardiffeos}
  \mathrm{Diffeomorphisms:} \quad \delta_{\xi} e^{I}= {\mathcal L}_{\xi} e^{I},
  \quad  \delta_{\xi}\omega^{IJ}= {\mathcal L}_{\xi} \omega^{IJ},
\end{eqnarray}
where $\tau^{IJ} (=-\tau^{JI})$ and ${\mathcal L}_{\xi}$ is the Lie derivative along the vector field $\rho = \xi^{I}\partial_{I}$, with $\partial_{I}$ being the dual basis of the frame $e^{I}$ \cite{Castillobook}. The transformations given in \eref{eq:varLorentzgroup} and \eref{eq:vardiffeos} are symmetries of Lovelock action since each term $\int a_{p} L_{n}^{p}$ is invariant under \eref{eq:varLorentzgroup} and quasi-invariant under \eref{eq:vardiffeos}. The converse of Noether's second theorem states that we can uncover a gauge symmetry from an identity relating the variational derivatives of the action. In the following section we construct these relations, referred to as `Noether identities'.

\section{Usual gauge symmetries from the converse of Noether's second theorem}\label{sec:Gauge}
We must emphasize that, as we are interested in analyzing the symmetries of Lovelock theory using the converse of Noether's second theorem, our approach is off-shell because such a hypothesis is demanded by the theorem. Furthermore,  we do not use the additional constriction $De^{I}=0$ at any point of this paper, which is usually assumed in the literature on the subject.

\subsection{Local Lorentz transformations }\label{subsec:Lorentz}
Although we already know that Lovelock action is invariant under local Lorentz transformations, it is instructive to review how this symmetry emerges from the application of the converse of Noether's second theorem.

In order to illustrate our method, we obtain step-by-step the Noether identity related to local Lorentz transformations for the first term of the Lovelock action \eref{eq:lovelockaction}, which is 
\begin{equation}
  \label{eq:Lovelockactionterm0}
  \int  L_{n}^{0} = \kappa \int
  \epsilon_{I_{1} \cdots I_{n}} e^{I_{1}}\wedge\cdots\wedge e^{I_{n}}.
\end{equation} 
This term is independent of $\omega^{IJ}$. Therefore, its only nonvanishing variational derivative is
\begin{equation}
  \label{eq:eulerep2}
  \mathcal{E}_{I}^{0}  =
  \left(-1\right)^{n-1} \kappa n \epsilon_{I I_{2}\cdots I_{n}}
  e^{I_{2}} \wedge \cdots \wedge e^{I_{n}}.
\end{equation}
We compute the wedge product of $e^{I}$ with $\mathcal{E}_{J}^{0}$, obtaining 
\begin{equation}
  \label{eq:euler1p0}
  e^{I} \wedge \mathcal{E}^{0}_{J} =
  \left(-1\right)^{n-1}
  \kappa n \epsilon_{J I_{2}\cdots I_{n}} e^{I} \wedge
  e^{I_{2}} \wedge \cdots \wedge e^{I_{n}}.
\end{equation}
Then, lowering the index $I$ and antisymmetrizing with respect to the indices $I,J$ in last expression, we obtain the Noether identity
\begin{eqnarray}
  \label{eq:NoetheridLorentzp0}
  e_{[I}\wedge \mathcal{E}^{0}_{J]}=0.
\end{eqnarray}
Analogously, we can obtain a Noether identity involving the variational derivatives of every term of the Lovelock action \eref{eq:lovelockaction} with $p$ curvature factors $\int L_{n}^{p}$.This variational derivatives are given in \eref{eq:eulerep} and \eref{eq:eulerwp}. By computing the covariant derivative of \eref{eq:eulerwp} and after a few algebra, we get the Noether identity
\begin{equation}
  D\mathcal{E}^{p}_{I J}  - e_{\left[ I \right.} \wedge
  \mathcal{E}^{p}_{\left. J \right]} =0. \label{eq:NoetheridLorentzp}
\end{equation}

Adding the Noether identities \eref{eq:NoetheridLorentzp0} and \eref{eq:NoetheridLorentzp} multiplied by the corresponding coefficient $a_{p}$ leads to a Noether identity involving the full variational derivatives $\mathcal{E}_{I}$ and $\mathcal{E}_{IJ}$, which is
\begin{eqnarray}
  \sum_{p=1}^{\left[ \case{n-1}{2}\right]} a_{p} D\mathcal{E}^{p}_{I J}
  - \sum_{p=0}^{\left[ \case{n-1}{2}\right]} a_{p} e_{\left[ I \right.} \wedge
  \mathcal{E}^{p}_{\left. J \right]}
  = D\mathcal{E}_{I J} - e_{\left[ I \right.} \wedge
  \mathcal{E}_{\left. J \right]}=0. \label{eq:NoetheridLorentz}
\end{eqnarray}
Multiplying this expression by arbitrary local parameters $\tau^{IJ} (=-\tau^{JI})$, we get the off-shell identity
\begin{equation}
  \label{eq:off-shellLorentz}
  \mathcal{E}_{I} \wedge
  \underbrace{\tau^{I}\,_{J}e^{J}}_{\delta_{\tau} e^{I}}
  + \mathcal{E}_{IJ} \wedge
  \underbrace{(-D\tau^{IJ})}_{\delta_{\tau}\omega^{IJ}}
  + \left(-1\right)^{n-1}
  d\left( \tau^{IJ}\mathcal{E}_{IJ} \right) = 0.  
\end{equation}

According to the converse of Noether's second theorem, we can read off a symmetry of the Lovelock action \eref{eq:lovelockaction} from the terms accompanying the variational derivatives $\mathcal{E}_{I}^{p}$ and $\mathcal{E}^{p}_{IJ}$ in the off-shell identity \eref{eq:off-shellLorentz}. We see that the gauge transformation that can be read off from \eref{eq:off-shellLorentz} is the infinitesimal local Lorentz transformation \eref{eq:varLorentzgroup}. Therefore, we have obtained local Lorentz symmetry by applying the converse of Noether's second theorem to the Lovelock action \eref{eq:lovelockaction}. Furthermore, we can observe from the Noether identities \eref{eq:NoetheridLorentzp0} and \eref{eq:NoetheridLorentzp} that every term $\int L_{n}^{p}$ of the Lovelock action \eref{eq:lovelockaction} is invariant under local Lorentz transformations\footnote{From the Lovelock action \eref{eq:lovelockaction} we observe that in even dimensions, there is a term that we have not analyzed yet, which is 
\begin{equation*}
  \label{eq:Eulerform2}
  \int L_{n}^{n/2} = \kappa  \int
  \epsilon_{I_{1}I_{2} \cdots I_{n-1}I_{n}} R^{I_{1}I_{2}}
  \wedge \cdots \wedge R^{I_{n-1}I_{n}},  
\end{equation*}
but, as we pointed out before, this term is topological, and therefore it is invariant under local Lorentz transformations. Because this term is topological, it will not be analyzed in the rest of this paper.}.

\subsection{Diffeomorphisms}\label{subsec:Diffeomorphisms}
Now we show that the symmetry under diffeomorphisms of the Lovelock action \eref{eq:lovelockaction} can also be uncovered by means of the converse of Noether's second theorem. As for the local Lorentz symmetry, we begin by analyzing the first term of the Lovelock action \eref{eq:lovelockaction}, which is
\begin{equation}
  \label{eq:lovelockp0indiffeos}
  \int L_{n}^{0} = \kappa \int \epsilon_{I_{1} \cdots I_{n}} e^{I_{1}} \wedge \cdots \wedge e^{I_{n}}.   
\end{equation}
First, we recall that the variational derivative of \eref{eq:lovelockp0indiffeos} with respect to the frame $e^{I}$ is
\begin{equation}
  \label{eq:eulere0indiffeos}
  \mathcal{E}_{I}^{0} = (-1)^{n-1} \kappa n \epsilon_{I I_{2} \cdots I_{n}} e^{I_{2}} \wedge \cdots \wedge e^{I_{n}}.   
\end{equation}
Then, we compute the covariant derivative $D\mathcal{E}^{0}_{I}$, obtaining
\begin{eqnarray}
  D  \mathcal{E}_{I}^{0} 
  & = (-1)^{n-1} \kappa n (n-1) \epsilon_{I I_{2} \cdots _{I_{n}}} De^{I_{2}} \wedge \cdots \wedge e^{I_{n}}
    \nonumber
  \\
  & = \mathcal{T}^{K}\,_{I J}e^{J}\wedge \mathcal{E}_{K}^{0},
    \label{eq:Deuler1p0}
\end{eqnarray}
with $\mathcal{T}^{I}\,_{JK}$ being the components of $De^{K}: = (1/2)\mathcal{T}^{K}\,_{IJ} e^{I}\wedge e^{J}$. Handling \eref{eq:Deuler1p0} we arrive at the Noether identity
\begin{eqnarray}
  \label{eq:Noetheriddiffeosp0}
  D\mathcal{E}^{0}_{I} =\left(\partial_{I} \intprod De^{K} \right) \wedge \mathcal{E}^{0}_{K},
\end{eqnarray}
where we have substituted
$\mathcal{T}^{K}\,_{IJ}e^{J} = \partial_{I} \intprod De^{K}$\footnote{The symbol `$\intprod$' stands for the contraction of a vector field $X$ and a $k$-form $\alpha$ yielding a $(k-1)$-form $X\intprod\alpha$ defined by
\begin{eqnarray}
 (X \intprod \alpha)(v_{1},\ldots,v_{k-1}):=k\alpha(X,v_{1},\ldots,v_{k-1}),\nonumber
\end{eqnarray}
where $v_{1}, \ldots ,v_{k-1}$ are vector fields \cite{Castillobook}.}.

The Noether identity \eref{eq:Noetheriddiffeosp0} suggests that we can relate $\mathcal{E}_{I}^{p}$ with its covariant derivative $D\mathcal{E}_{I}^{p}$ (and possibly with $\mathcal{E}_{IJ}^{p}$). This results to be true, and following the same procedure that leads us to the Noether identity \eref{eq:Noetheriddiffeosp0}, we can construct the Noether identity
\begin{eqnarray}
  \label{eq:Noetheriddiffeosp}
  D\mathcal{E}^{p}_{I} = \left(\partial_{I} \intprod De^{K}\right) \wedge \mathcal{E}^{p}_{K} + \left(\partial_{I} \intprod R^{KL} \right) \wedge \mathcal{E}^{p}_{KL},
\end{eqnarray}
which is valid for all the other terms $ \int L_{n}^{p}$ of the Lovelock action \eref{eq:lovelockaction}\footnote{Recall that $p$ counts the number of curvature factors in $L_{n}^{p}$, and also that $\int L_{n}^{n/2}$ is topological for even $n$. So, there is no need to construct a Noether identity for this last term.}. However, we must analyze separately the term $\int L_{n}^{(n-1)/2}$ with odd $n$, because the covariant derivative of $\mathcal{E}_{I}^{(n-1)/2}$ vanishes, \textit{i.e.} 
\begin{equation}
  \label{eq:Deuler1vanishingodd}
  D\mathcal{E}_{I}^{(n-1)/2} = 0,
\end{equation}
which emerges using the Bianchi's identity $DR^{IJ}=0$. In this last case the Noether identity \eref{eq:Noetheriddiffeosp} can be written as a linear combination of other two independent Noether identities.

The first Noether identity involving the variational derivatives of $\int L_{n}^{(n-1)/2}$ is \eref{eq:Deuler1vanishingodd}. We will analyze the symmetry that emerges from \eref{eq:Deuler1vanishingodd} when we study the so-called `local translations' in subsection \ref{subsec:local0}, but for now let us keep looking for the Noether identity which leads to diffeomorphisms.

The second Noether identity involving the variational derivatives of $\int L_{n}^{(n-1)/2}$ is
\begin{equation}
  \label{eq:Noetheridnew(n-1)/2}
  \left ( \partial_{I} \intprod D e^K \right )
  \wedge \mathcal{E}^{(n-1)/2}_{K} 
  + \left( \partial_I \intprod R^{KL} \right)
  \wedge \mathcal{E}^{(n-1)/2}_{KL} =0, 
\end{equation}
which can be verified directly from \eref{eq:eulerep} and \eref{eq:eulerwp}. This Noether identity had not been reported in the literature and, according to the converse of Noether's second theorem, it reveals a new gauge symmetry of the action $\int L_{n}^{(n-1)/2}$. This fact requires further analysis. So, we devote section \ref{sec:newsymmetry} to the analysis of the symmetries of the last term of the Lovelock action \eref{eq:lovelockaction} in odd dimensions.

Coming back to the diffeomorphism symmetry, we can construct from \eref{eq:Deuler1vanishingodd} and \eref{eq:Noetheridnew(n-1)/2} a Noether identity analogous to \eref{eq:Noetheriddiffeosp}, which is
\begin{equation}
  D\mathcal{E}_{I}^{(n-1)/2} = \left ( \partial_{I} \intprod D e^K \right ) \wedge\mathcal{E}^{(n-1)/2}_{K} + \left( \partial_I \intprod R^{KL} \right) \wedge \mathcal{E}^{(n-1)/2}_{KL}.
  \label{eq:Noetheriddiffeos(n-1)/2}
\end{equation}
Therefore, as we did for local Lorentz symmetry, from the Noether identities \eref{eq:Noetheriddiffeosp0}, \eref{eq:Noetheriddiffeosp} and \eref{eq:Noetheriddiffeos(n-1)/2} we construct a Noether identity involving the variational derivatives of the full Lovelock action \eref{eq:lovelockaction} $\mathcal{E}_{I}$ and $\mathcal{E}_{IJ}$, namely
\begin{eqnarray}
  \fl \sum_{p=0}^{\left[\case{n-1}{2}\right]} a_{p}D\mathcal{E}^{p}_{I} -
  \left ( \partial_{I} \intprod D e^K \right )
  \wedge \sum_{p=0}^{\left[\case{n-1}{2}\right]} a_{p} \mathcal{E}^{p}_{K} 
  - \left( \partial_I \intprod R^{KL} \right)
  \wedge \sum_{p=1}^{\left[\case{n-1}{2}\right]} a_{p}\mathcal{E}^{p}_{KL} 
  \nonumber
  \\
   \fl = D\mathcal{E}_{I} - \left ( \partial_{I} \intprod D e^K \right )
  \wedge \mathcal{E}_{K} 
  - \left( \partial_I \intprod R^{KL} \right)
  \wedge \mathcal{E}_{KL}=0.
\label{eq:Noetheriddiffeos}
\end{eqnarray}
Multiplying \eref{eq:Noetheriddiffeos} by the arbitrary local parameter $\chi^I$ and after a few algebra, we get the off-shell identity
\begin{eqnarray}
  \label{eq:off-shelldiffeos}
  \mathcal{E}_{I}
  \wedge \underbrace{\left( D \chi^{I} +
  \chi \intprod De^{I} \right)}_{\delta_{\chi} e^{I}} +
  \mathcal{E}_{IJ} \wedge
  \underbrace{ \left(\chi \intprod R^{IJ}\right)}_{\delta_{\chi}
  \omega^{IJ}}
  +\left(-1\right)^{n}d\left(\chi^{I}\mathcal{E}_{I}\right)=0,
\end{eqnarray}
with $\chi:=\chi^{I}\partial_{I}$. According to the converse of Noether's second theorem the gauge symmetry of the Lovelock action \eref{eq:lovelockaction} involved in \eref{eq:off-shelldiffeos} is given by the terms accompanying the variational derivatives $\mathcal{E}_{I}$ and $\mathcal{E}_{IJ}$. In order to appreciate its meaning, we rewrite the gauge transformation involved in \eref{eq:off-shelldiffeos} as
\begin{eqnarray}
  \fl \delta_{\chi} e^I
  &= {\mathcal L}_{\chi} e^{I} + \tau^{I}\,_{J}e^{J} = \left(\delta_{\xi}+\delta_{\tau}\right)e^{I},
  &\quad \left(\tau^{IJ}:= \rho \intprod \omega^{IJ} , \quad\xi^{I} := \chi^{I}\right), 
  \label{eq:Kiriushchevatodiffeose}
  \\
  \fl \delta_{\chi} \omega^{IJ} 
  &= \mathcal{L}_{\chi} \omega^{IJ}-D\tau^{IJ}=\left(\delta_{\xi}+\delta_{\tau}\right)\omega^{IJ}.
  &  
  \label{eq:Kiriushchevatodiffeosw}
\end{eqnarray}
Hence, we observe that the symmetry involved in \eref{eq:off-shelldiffeos} is nothing else than an infinitesimal diffeomorphism \eref{eq:vardiffeos} plus a local Lorentz transformation \eref{eq:varLorentzgroup} with field-dependent gauge parameters, which is sometimes referred to as an `improved diffeomorphism' in the context of the Palatini action \cite{RevModPhys.48.393}.

In this way, we have proved that the Lovelock action \eref{eq:lovelockaction} is diffeomorphism invariant by means of the converse of Noether's second theorem. Furthermore, from the Noether identities \eref{eq:Noetheriddiffeosp0}, \eref{eq:Noetheriddiffeosp} and \eref{eq:Noetheriddiffeos(n-1)/2} we can observe that each term $\int L_{n}^{p}$ is also invariant under diffeomorphisms.

\subsection{Three-dimensional local translations}\label{subsec:3dLocaltranslations}
As we show below, Lovelock theory \eref{eq:lovelockaction} coincides with general relativity in three dimensions. It is very well-known that the gauge freedom of three-dimensional general relativity can be described by local Lorentz transformations and diffeomorphisms, or equivalently by local
Lorentz transformations and a symmetry referred to as `local translations' \cite{Witten:1988hc,ACHUCARRO198689} (see also \cite{carlip2003quantum}). In order to generalize three-dimensional local translations of general relativity to the Lovelock action \eref{eq:lovelockaction} from the perspective of the converse of Noether's second theorem, we gain some insight first by reviewing how this symmetry emerges in the three-dimensional case \cite{Montesinos2017, Montesinos2018}.

In three dimensions the Lovelock action (\ref{eq:lovelockaction}) acquires the form
\begin{equation}
  \label{eq:lovelockaction3d}
  S_{3}[e,\omega] = \kappa \int \left(a_{0}\epsilon_{I_{1} I_{2} I_{3}}
    e^{I_{1}} \wedge e^{I_{2}} \wedge e^{I_{3}} + a_{1} \epsilon_{I_{1} I_{2} I_{3}}R^{I_{1}I_{2}} \wedge e^{I_{3}} \right),  
\end{equation}
and its variational derivatives are
\begin{eqnarray}\label{eq:eulere3d}
  \mathcal{E}_{I} = a_{0}\mathcal{E}_{I}^{0}
  + a_{1}\mathcal{E}_{I}^{1} =
  3  a_{0} \kappa  \epsilon_{I I_{2} I_{3}}
  e^{I_{2}} \wedge e^{I_{3}}
  +  a_{1} \kappa  \epsilon_{I I_{2} I_{3}}
  R^{I_{2}I_{3}},
  \\\label{eq:eulerw3d}
  \mathcal{E}_{IJ} = a_{1}\mathcal{E}_{IJ}^{1}
  = a_{1} \kappa \epsilon_{IJ I_{3}}De^{I_{3}}.
\end{eqnarray}
Computing the exterior covariant derivative of \eref{eq:eulere3d} we obtain
\begin{eqnarray}
  \fl  D\mathcal{E}_{I} 
  & = a_{0} D\mathcal{E}_{I}^{0} + a_{1} D\mathcal{E}_{I}^{1} = -3! a_{0} \kappa \epsilon_{I J I_{3}} e^{J} \wedge De^{I_{3}} + a_{1} \kappa \epsilon_{I I_{2} I_{3}} DR^{I_{2}I_{3}}.
  \label{eq:Deulere3d}
\end{eqnarray}
From Bianchi's identity $DR^{IJ}=0$ we see that $D\mathcal{E}_{I}^{1}$ vanishes. Hence, handling \eref{eq:Deulere3d} we arrive at the Noether identity
\begin{equation}
  \label{eq:Noetherlocaltrans3d}
  D\mathcal{E}_{I}= 3!(a_{0}/a_{1})e^{J} \wedge \mathcal{E}_{JI}.
\end{equation}
In the standard treatment of three-dimensional local translations one usually sets $ 3!(a_{0}/a_{1}) = -2 \Lambda$, where $\Lambda$ is the
cosmological constant \cite{carlip2003quantum}. Therefore, the Noether identity\footnote{See also \cite{Montesinos2018Geo} to see the Noether
  identity in the case of Witten's exotic action for three-dimensional gravity with cosmological constant.} \eref{eq:Noetherlocaltrans3d} reads \cite{Montesinos2017}
\begin{equation}
  \label{eq:Noetherlocaltrans3dlambda}
  D\mathcal{E}_{I}-2\Lambda e^{J}\wedge \mathcal{E}_{IJ} = 0.
\end{equation}
Multiplying \eref{eq:Noetherlocaltrans3dlambda} by the arbitrary local parameter $\rho^{I}$ and manipulating the resulting expression we get the off-shell identity
\begin{equation}
  \mathcal{E}_{I} \wedge \underbrace{ D \rho^{I}}_{\delta_{\rho} e^{I}} + \mathcal{E}_{IJ} \wedge \underbrace{ 2 \Lambda \rho^{[I}e^{J]}}_{\delta_{\rho} \omega^{IJ}} + d\left(-\rho^{I}\mathcal{E}_{I}\right)=0.
\end{equation}
Resorting to the converse of Noether's second theorem, the quantities which appear in this last expression multiplying the variational derivatives are the transformations associated to a gauge symmetry; in this case, the so-called `local translations'.

We notice three important facts from \eref{eq:Noetherlocaltrans3d} and \eref{eq:Noetherlocaltrans3dlambda}:
 
a) If $\Lambda \neq 0$ all coefficients $a_{p}$ of the three-dimensional Lovelock action must be nonvanishing.
 
b) If $\Lambda \neq 0$ the Noether identity from which local translations symmetry emerges relates consecutive coefficients of the action.
   
c) If $\Lambda = 0$ the only one nonvanishing coefficient of the three-dimensional Lovelock action is $a_{1}$.

\subsection{Local translations with $\Lambda \neq 0$}\label{subsec:localno0}
In this subsection we prove that three-dimensional local translations with $\Lambda \neq 0$ can be generalized to the $n$-dimensional Lovelock action \eref{eq:lovelockaction} for a particular choice of the coefficients $a_{p}$ only. This symmetry has been already known for a long time. For instance, it has been obtained in \cite{Zanelli2012} with a method that relies on $(2n+2)$-dimensional Chern-Simons forms. Here, on the other hand, we uncover it simply using the converse of Noether's second theorem and then our approach is conceptually simpler and more elegant than previous ones because it relies on a fundamental principle of theoretical physics. Moreover, the approach of this paper has been successfully applied to general relativity with cosmological constant with and without matter fields \cite{Montesinos2017, Montesinos2018}.  

In order to obtain this symmetry, we consider a generic term $\int L^{p}_{n}$ of the Lovelock action \eref{eq:lovelockaction}, whose variational derivatives are given in \eref{eq:eulerep} and \eref{eq:eulerwp}.

As the item b) of subsection \ref{subsec:3dLocaltranslations} suggests, we compute
\begin{eqnarray}
  \fl D \mathcal{E}_{I}^{p} 
  & = & \left(-1\right)^{n-1} 
        \kappa \left(n-2p\right) \left(n-2p-1\right)
        \epsilon_{I I_{2} I_{3} \cdots I_{2p}I_{2p+1}I_{2p+2}I_{2p+3}  \cdots I_{n}}  
        \nonumber\\
  \fl                    
  &   & \times R^{I_{2}I_{3}} \wedge \cdots \wedge R^{I_{2p}I_{2p+1}} \wedge
        De^{I_{2p+2}} \wedge e^{I_{2p+3}}\wedge \cdots \wedge e^{I_{n}},
        \label{eq:Deulerepinlocal}
\end{eqnarray}
for $0 \leq p \leq \left[(n-3)/2\right]$. This expression can be rewritten as
\begin{equation}
  D\mathcal{E}^{p}_{I} = \case{\left(n-2p\right) \left(n-2p-1\right)}{\left(p+1\right) \left(n-2p-2\right)}e^{J} \wedge \mathcal{E}^{p+1}_{J I},
  \quad $for $0 \leq p \leq \left[(n-3)/2\right].
  \label{eq:relationDeuler1euler2contracted}
\end{equation}
Notice that equation \eref{eq:relationDeuler1euler2contracted} does not involve the variational derivatives of all the terms of the Lovelock action, leading to the following facts:
 
a) For odd $n$, equation \eref{eq:relationDeuler1euler2contracted} does not give us a relationship for $D\mathcal{E}^{(n-1)/2}$. However, we have seen
in subsection \ref{subsec:Diffeomorphisms} that this term vanishes. This fact will allow us to obtain local translations with $\Lambda \neq 0$ in odd dimensions.
   
b) For even $n$, equation \eref{eq:relationDeuler1euler2contracted} gives no a relationship for $D\mathcal{E}^{(n-2)/2}$. As we will see below, this term prevents us from obtaining local translations in the even-dimensional case.
   
c) Also for even $n$, $D\mathcal{E}^{n/2}_{I}$ is not involved in \eref{eq:relationDeuler1euler2contracted}.  However, $\int L^{n/2}_{n}$ is topological, which means that $\mathcal{E}^{n/2}_{I}$ (and consequently $D\mathcal{E}^{n/2}_{I}$) trivially vanishes.
 
From these observations, we conclude that we must consider the even- and odd-dimensional cases separately:
 
(i) Odd-dimensional manifolds $\mathcal{M}^{n}$. Consider the Lovelock action \eref{eq:lovelockaction} whose variational derivatives $\mathcal{E}_{I}$ and $\mathcal{E}_{IJ}$ are given in \eref{eq:eulere} and \eref{eq:eulerw}, respectively. Computing the covariant derivative of $\mathcal{E}_{I}$, and recalling that $D\mathcal{E}_{I}^{(n-1)/2}=0$, we get
\begin{equation}
  \label{eq:euler1totalassum}
  D\mathcal{E}_{I} = \sum^{\case{n-3}{2}}_{p=0}a_{p}D\mathcal{E}_{I}^{p}.
\end{equation}
Substituting \eref{eq:relationDeuler1euler2contracted} into the right-hand side of \eref{eq:euler1totalassum} we get
\begin{eqnarray}
  \label{eq:otra}
  D\mathcal{E}_{I} =
  \sum^{\case{n-1}{2}}_{p=1}
  a_{p-1}\case{(n-2p+2)(n-2p+1)}{p(n-2p)} e^{J} \wedge
  \mathcal{E}_{JI}^{p}.
\end{eqnarray}
In order to relate \eref{eq:otra} with $e^{J}\wedge\mathcal{E}_{JI}$, we ask the coefficients $a_{p}$ to satisfy the relationship
\begin{equation}
  \label{eq:conditionrecurrence}
  2a_{p}\Lambda=-a_{p-1}\case{(n-2p+2)(n-2p+1)}{p(n-2p)},
  \quad $for  $p=1,\ldots,\case{n-1}{2}, 
\end{equation}
which finally allows us to obtain the Noether identity
\begin{equation}
  \label{eq:off-shelldeSitter}
  D\mathcal{E}_{I} -2 \Lambda e^{J} \wedge \mathcal{E}_{IJ}=0,
  \quad \Lambda \neq 0.
\end{equation}
Note that this Noether identity is identical to the one for the three-dimensional case. Nevertheless, if we look carefully at the action principle we realize that $\Lambda$ is not the constant accompanying the volume term as it happens in the three-dimensions. The definition of $\Lambda$ is given by \eref{eq:conditionrecurrence}. Multiplying \eref{eq:off-shelldeSitter} by the real local parameters $\rho^{I}$ and handling the resulting expression, the off-shell identity emerges
\begin{equation}
  \label{eq:NoetheriddeSitter}
  \mathcal{E}_{I} \wedge \underbrace{D\rho^{I}}_{\delta_{\rho} e^{I}} + \mathcal{E}_{IJ} \wedge \underbrace{2\Lambda \rho^{[I}e^{J]}}_{\delta_{\rho} \omega^{IJ}} +\left(-1\right)^{n}d\left(\rho^{I}\mathcal{E}_{I}\right) = 0,
\end{equation}
from which, according to the converse of Noether's second theorem, the local translations can be read off
\begin{eqnarray}
  \label{eq:varedeSitter}
  \delta e^{I} &= & D\rho^{I},
  \\
  \label{eq:varwdeSitter}
  \delta \omega^{IJ} &= & 2 \Lambda \rho^{[I} e^{J]}.
\end{eqnarray}
Coming back to the recurrence relation \eref{eq:conditionrecurrence}, we obtain a solution by handling the relationship among the coefficients $a_{p}$ as follows
\begin{eqnarray}
  \nonumber
  \fl \frac{a_{p}}{a_{p-1}}
  &=&-\frac{(n-2p+2)(n-2p+1)}{2 \Lambda p(n-2p)}
  \\
  \nonumber
  \fl & =& \left(\frac{(-1)^{p-1}}{(-1)^{p}}\right)
        \left(\frac{\Lambda^{p-1}}{\Lambda^{p}}\right)
        \left(\frac{(p-1)}{p!}\right)
        \left(\frac{(n-2(p-1))}{(n-2p)}\right)
        \left(\frac{n-1}{2}-(p-1)\right)
  \\
  \fl & =& \left(\frac{(-1)^{p-1}}{(-1)^{p}}\right)
        \left(\frac{\Lambda^{p-1}}{\Lambda^{p}}\right)
        \left(\frac{(p-1)!}{p!}\right)
        \left(\frac{(n-2(p-1))}{(n-2p)}\right)
        \left(\frac{\alpha}{\alpha}\right)\nonumber
  \\
  \label{eq:howsolutioncoeficients}  
       \fl &&\times \frac{\left(\case{n-1}{2}-(p-1)\right)!}
        {\left(\case{n-1}{2}-p\right)!},
\end{eqnarray}
where we have introduced a nonvanishing constant $\alpha$. From \eref{eq:howsolutioncoeficients} we can read off the form of $a_{p}$ and $a_{p-1}$ by associating the quantities where $p$ appears to $a_{p}$ and the quantities where $p-1$ appears to $a_{p-1}$. In order to get rid of $\alpha$, we note from the expression for $a_p$ 
that making $p=0$ we get $\alpha= n \left(\case{n-1}{2}\right)! a_0$, which leads to the final expression for the coefficients
\begin{equation}
  \label{eq:sequencesolution}
  a_{p} = \frac{n\left(-1\right)^{p}\left(\case{n-1}{2}\right)!}
  {\Lambda^{p}\left(n-2p\right)p!\left(\case{n-1}{2}-p\right)!}
  a_{0}.
\end{equation}
From \eref{eq:sequencesolution} we can observe that for local translations with $\Lambda \neq 0$ become a symmetry of the Lovelock action \eref{eq:lovelockaction} in odd dimensions all the coefficients $a_{p}$ must be nonvanishing.  Note also that the coefficients $a_{p}$ that satisfy the condition \eref{eq:sequencesolution} depend on the dimension $n$ of the manifold, while the gauge transformation \eref{eq:varedeSitter} and \eref{eq:varwdeSitter} is independent of $n$.

Finally, it has been already proved in \cite{Mansouri1977} (see also \cite{Zanelli2012, MontesinosBF}) that the symmetry \eref{eq:varedeSitter}-\eref{eq:varwdeSitter} together with local Lorentz transformations correspond to the de Sitter group $SO(1,n)$ ($\Lambda > 0$) or the anti-de Sitter group $SO(2,n-1)$ ($\Lambda < 0$) \cite{carlip2003quantum, Zanelli2012}, considering  a larger gauge connection built up from both the $SO(\sigma)$-connection $\omega^{IJ}$ and the frame $e^{I}$. On the other hand, for Euclidean manifolds ($\sigma = +1$) the symmetry \eref{eq:varedeSitter}-\eref{eq:varwdeSitter} together with the $SO(n)$ transformations correspond to the $SO(n+1)$ group ($\Lambda > 0$) and to the $SO(1,n)$ group ($\Lambda < 0$). The transformation of the Lovelock action under the (anti-)de Sitter group has also been studied in \cite{Zanelli2012} in the context of Chern-Simons forms in gravitational theories (see also \cite{Troncoso2000}). For the rest of this work, when we talk about `the (anti-)de Sitter group' we are referring to the $SO(p,q)$ group associated with the signature of the frame rotation group. 

(ii) Even-dimensional manifolds $\mathcal{M}^{n}$. A completely analogous procedure can be used in the even-dimensional case to obtain the relationship
\begin{equation}
  \label{eq:NoetheriddeSitternot}
  D\mathcal{E}_{I} = -2 \Lambda e^{J} \wedge
  \mathcal{E}_{JI} + a_{(n-2)/2}D\mathcal{E}^{(n-2)/2}_{I}, 
\end{equation}
where
\begin{equation}
  \label{eq:sequencesolutioneven}
  a_{p}=\frac{(-1)^{p}n!
  \left(\case{n-2}{2}-p\right)!}
  {4^{p}\left(n-2p\right)!
  \Lambda^{p}p!\left(\case{n-2}{2}\right)!}a_{0},
  \quad $for$\,\, p=0,...,(n-2)/2.
\end{equation}
We observe that \eref{eq:NoetheriddeSitternot} is not generically of the form of a Noether identity, because it has the extra term
\begin{equation}
  \label{eq:Extratermeven}
  a_{(n-2)/2} D\mathcal{E}^{(n-2)/2}_{I}, 
\end{equation}
with
\begin{equation}
  \label{eq:non-vaninshingterm}
  D\mathcal{E}^{(n-2)/2}_{I} =
  \left(-1\right)^{n-1}2\kappa
  \epsilon_{I I_{2} \cdots I_{n}}R^{I_{1} I_{2}} \wedge
  R^{I_{n-2} I_{n-1}} \wedge De^{I_{n}}.
\end{equation}
Since \eref{eq:Extratermeven} is not a total differential nor can be generically rewritten in terms of $\mathcal{E}_{I}$ and $\mathcal{E}_{IJ}$, we conclude that local translations with $\Lambda \neq 0$ given in \eref{eq:varedeSitter}-\eref{eq:varwdeSitter} are not a symmetry of the Lovelock action \eref{eq:lovelockaction} in even-dimensions, except for the case where the action is given by the Euler term. However, as was reported in \cite{Montesinos2017}, Noether identities different from \eref{eq:off-shelldeSitter} can be obtained for particular cases of the Lovelock action (for instance for the Palatini action studied in that work), leading to different generalizations of local translations which are valid in even dimensions.

\subsection{Local translations with $\Lambda=0$}\label{subsec:local0}
Analogously as we did for local translations with $\Lambda \neq 0$, we split the analysis in odd- and even-dimensional cases:
   
(i) Odd-dimensional manifolds $\mathcal{M}^{n}$. Consider the single-term Lovelock action $\int L_{n}^{(n-1)/2}$. We have seen in subsection \ref{subsec:Diffeomorphisms} that \eref{eq:Deuler1vanishingodd} is a Noether identity by itself. Therefore, multiplying \eref{eq:Deuler1vanishingodd} by arbitrary local parameters $\rho^{I}$ and manipulating the resulting expression we arrive at the off-shell identity 
\begin{equation}
  \label{eq:off-shellPoincare}
  \mathcal{E}^{(n-1)/2}_{I} \wedge
  \underbrace{D\rho^{I}}_{\delta_{\rho} e^{I}} +
  d\left(-\rho^{I}\mathcal{E}^{(n-1)/2}_{I}\right) = 0.
\end{equation}
Resorting to the converse of Noether's second theorem again, we observe that local translations with $\Lambda=0$ emerge from the quantities accompanying the variational derivative $\mathcal{E}^{(n-1)}_{I}$ in \eref{eq:off-shellPoincare}, namely
\begin{eqnarray}
  \label{eq:varePoincare}
  \delta e^{I} = & D\rho^{I},\\
  \label{eq:varwPoincare}
  \delta \omega^{IJ} = & 0.
\end{eqnarray}
Local translations given by equations \eref{eq:varePoincare}-\eref{eq:varwPoincare} together with local Lorentz transformations form the Poincar\'e group if $\sigma=-1$ and the Euclidean group if $\sigma=1$.

(ii) Even-dimensional manifolds $\mathcal{M}^{n}$. There is no term of the Lovelock action \eref{eq:lovelockaction} whose variational derivative fulfills
\begin{equation}
  D\mathcal{E}^{p}_{I} = 0,
\end{equation}
except for the topological term
\begin{equation}
  \label{eq:TopologicalLovelock}
  S_{n}[e,\omega]= \kappa \int{a_{n/2}\epsilon_{I_{1}I_{2} \cdots I_{n-1} I_{n}}}
  R^{I_{1}I_{2}}\wedge \cdots \wedge R^{I_{n-1}I_{n}},
\end{equation}
as one can see from \eref{eq:Deulerepinlocal}.  Therefore, we can not construct the Noether identity associated to the local translations with $\Lambda=0$ in the even-dimensional case analogously as we did in the odd-dimensional one. As far as we know, there is no symmetry analogous to \eref{eq:varePoincare}-\eref{eq:varwPoincare} for the Lovelock action \eref{eq:lovelockaction} in even dimensions.

\section{A new gauge symmetry of Lovelock action if its coefficients satisfy \eref{eq:sequencesolution}}\label{subsec:NewsymmetryLnot0}
As we have seen in subsection \ref{subsec:localno0}, Lovelock action \eref{eq:lovelockaction} is quasi-invariant under local translations with $\Lambda \neq 0$ \eref{eq:varedeSitter}-\eref{eq:varwdeSitter} if the coefficients $a_{p}$ satisfy the relation \eref{eq:conditionrecurrence}. The Noether identity from which local translations come from is given by \eref{eq:off-shelldeSitter}. This identity can be combined with the Noether identity associated to `improved diffeomorphisms'  \eref{eq:Noetheriddiffeos}, which leads to following the Noether identity 
\begin{eqnarray}
  \left(\partial_{I} \intprod De^{K}\right) \wedge \mathcal{E}_{K}
  + \left[
  \left(\partial_{I} \intprod R^{KL}\right)-2 \Lambda \delta^{[K}_{I}\delta^{L]}_{J}e^{J}
  \right] \wedge \mathcal{E}_{KL}=0.
  \label{eq:Noetheridnuevalno0}
\end{eqnarray}
We multiply \eref{eq:Noetheridnuevalno0} by arbitrary local parameters $\varepsilon^{I}$, and after a few algebra we get the off-shell identity 
\begin{eqnarray}
  \mathcal{E}_{I} \wedge \underbrace{\left(\varepsilon \intprod De^{I}\right)}_{\delta_{\varepsilon} e^{I}}
  + \mathcal{E}_{IJ} \wedge \underbrace{\left[
  \left( \varepsilon \intprod R^{IJ}\right) - 2 \Lambda \varepsilon^{[I} e^{J]}
  \right]}_{\delta_{\varepsilon} \omega^{IJ}} =0,
  \label{eq:Off-shellnuevalno0}
\end{eqnarray}
with $\varepsilon = \varepsilon^I \partial_I$, from which the gauge symmetry can be read off
\begin{eqnarray}
  \delta_{\varepsilon} e^{I} 
  & = \varepsilon \intprod De^{I},
  \label{eq:varenuevaL}\\
  \delta_{\varepsilon} \omega^{IJ} 
  & = \varepsilon \intprod R^{IJ} - 2 \Lambda \varepsilon^{[I} e^{J]}.
  \label{eq:varwnuevaL}
\end{eqnarray}

Indeed, we observe that the Lovelock action \eref{eq:lovelockaction}, whose coefficients $a_{p}$ satisfy \eref{eq:sequencesolution}, is quasi-invariant under the gauge transformation \eref{eq:varenuevaL}-\eref{eq:varwnuevaL} because 
\begin{eqnarray}
  \label{eq:variationNewLnot0}
  \fl \delta S_{n}  
  = & -\kappa \int d \left( \sum_{p=1}^{(n-1)/2}
      a_{p} p \epsilon_{I J I_{3}I_{4}\cdots I_{2p-1}I_{2p}I_{2p+1}\cdots I_{n}}  
      \right. 
      \nonumber\\
  \fl 
    &  \Biggl. \times R^{I_{3}I_{4}} \wedge \cdots \wedge R^{I_{2p-1}I_{2p}} \wedge
      e^{I_{2p+1}} \wedge \cdots \wedge e^{I_{n}} \wedge \left(\varepsilon \intprod R^{IJ} - 2 \Lambda \varepsilon^{[I} e^{J]}\right) \Biggr).
\end{eqnarray}
Note that in three dimensions ($n=3$) the gauge symmetry \eref{eq:varenuevaL}-\eref{eq:varwnuevaL} becomes trivial because it is 
proportional to the variational derivatives \eref{eq:eulere3d} and \eref{eq:eulerw3d} \cite{Henneaux1994} 
\begin{eqnarray}
  \delta_{\varepsilon} e^{I} 
  & = -\frac{\sigma \Lambda }{3! \kappa a_{0} }\epsilon^{IJK} \left( \varepsilon \intprod \mathcal{E}_{JK}\right),
  \label{eq:varenuevatrivial}\\
  \delta_{\varepsilon} \omega^{IJ} 
  & = -\frac{ \sigma \Lambda }{3! \kappa a_{0} }\epsilon^{IJK} \left( \varepsilon \intprod \mathcal{E}_{K}\right).
  \label{eq:varwnuevatrivial}
\end{eqnarray}
However, if $n>3$ we observe from \eref{eq:eulere} and
\eref{eq:eulerw} that the new symmetry \eref{eq:varenuevaL}-\eref{eq:varwnuevaL} is \textit{not} trivial.

Therefore, if $\mathcal{M}^{n}$ is odd-dimensional with $n>3$ and the coefficients of Lovelock action satisfy \eref{eq:sequencesolution}, then the complete set of gauge symmetries of Lovelock action is composed of local Lorentz transformations ($\delta_{\tau}$), local translations with $\Lambda \neq 0$ \eref{eq:varedeSitter}-\eref{eq:varwdeSitter} ($\delta_{\rho}$), and also the new gauge symmetry \eref{eq:varenuevaL}-\eref{eq:varwnuevaL} ($\delta_{\varepsilon}$). The commutator algebra of this set, acting on both the frame and the connection, reads\footnote{To compute the algebra we assume in this section that the gauge parameters are field-independent. In contrast, in \ref{app:completealgebra} we compute the algebras of various equivalent sets of gauge symmetries of Lovelock action in the generic case, i.e, allowing the gauge parameters to depend on the fields.} 
\begin{eqnarray}
  \fl \left[\delta_{\tau_{1}},\delta_{\tau_{2}}\right] 
  & =\delta_{\tau_{3}}, 
  & \left(
    {\tau_{3}}^{IJ} := 2 {\tau_{1}}^{[ I \vert K} {\tau_{2}}^{\vert J]}{}_{K}
    \right),
    \label{eq:LorentzLorentz0}
  \\
  \fl \left[\delta_{\rho_{1}},\delta_{\rho_{2}}\right]
  & =\delta_{\tau}, 
  &  \left(
    \tau^{IJ} := 2\Lambda{\rho_{1}}^{[I}{\rho_{2}}^{J]}
    \right),
    \label{eq:transtrans0}
  \\
  \fl \left[\delta_{\tau} , \delta_{\rho}\right]
  & =\delta_{\rho_{1}}, 
  & \left(
    {\rho_{1}}^{I}:=-{\tau^{I}}_{J}\rho^{J}
    \right),
    \label{eq:Lorentztrans0}
  \\
  \fl \left[\delta_{\varepsilon_{1}},\delta_{\varepsilon_{2}}\right]
  & = \delta_{\tau} + \delta_{\rho} + \delta_{\varepsilon_{3}},
  &  \left( 
    \tau^{IJ} := \varepsilon_{2} \intprod \left( \varepsilon_{1} \intprod R^{IJ}\right) - 2 \Lambda {\varepsilon_{1}}^{[I} {\varepsilon_{2}}^{J]}\right., 
    \nonumber\\
  \fl
  &
  & \quad \rho^{I} := \varepsilon_{1}\intprod \left(\varepsilon_{2} \intprod De^{I} \right),
    \nonumber\\
  \fl
  &
  & \bigl.\quad {\varepsilon_{3}}^{I} := \varepsilon_{1} \intprod \left(\varepsilon_{2}\intprod De^{I}\right)
    + \varepsilon_{2} \intprod D {\varepsilon_{1}}^{I} - \varepsilon_{1} \intprod D {\varepsilon_{2}}^{I}\Bigr),
    \label{eq:NewLNewL0}
  \\
  \fl \left[\delta_{\tau},\delta_{\varepsilon}\right]
  & = \delta_{\varepsilon_{1}}, 
  & \left(
    {\varepsilon_{1}}^{I} :=-\tau^{I}{}_{J}\varepsilon^{J}
    \right),
    \label{eq:LorentzNewL0}
  \\
  \fl \left[\delta_{\rho},\delta_{\varepsilon}\right]
  & = \delta_{\varepsilon_{1}}, 
  & \left( 
    {\varepsilon_{1}}^{I} := -\varepsilon \intprod D\rho^{I} 
    \right).
    \label{eq:transNewL0}
\end{eqnarray}
We observe that the commutator algebra of the gauge symmetries is closed. The equations \eref{eq:LorentzLorentz0}-\eref{eq:Lorentztrans0} reflect the fact that the commutators of local Lorentz transformations and local translations with $\Lambda \neq 0$ form together the algebra of the de Sitter ($\Lambda >0$) or anti-de Sitter ($\Lambda <0$) group. On the other hand, \eref{eq:NewLNewL0} states that the commutator of two new symmetries \eref{eq:varenuevaL}-\eref{eq:varwnuevaL} is a linear combination of a local Lorentz transformation, a local translation and a transformation of the type \eref{eq:varenuevaL}-\eref{eq:varwnuevaL}, with field-dependent gauge parameters. Finally, the relationship \eref{eq:LorentzNewL0} (\eref{eq:transNewL0}, respectively) reveals that the commutator of a local Lorentz transformation (local translation) with the new gauge symmetry is again a transformation of the type of \eref{eq:varenuevaL}-\eref{eq:varwnuevaL}, with a rotated (translated) gauge parameter.

As we have seen in subsection \eref{subsec:Diffeomorphisms}, the Lovelock action \eref{eq:lovelockaction} is also quasi-invariant under diffeomorphisms. However, as the set of gauge symmetries formed by the local Lorentz transformations, local translations with $\Lambda \neq 0$ and the symmetry \eref{eq:varenuevaL}-\eref{eq:varwnuevaL} is complete, diffeomorphisms must be a linear combination of these transformations. A straightforward computation shows that in fact an infinitesimal diffeomorphism can be rewritten as
\begin{eqnarray}
  \delta_{\xi}e^{I}
  & ={\mathcal L}_{\xi }e^{I} 
  & = \left(\delta_{\tau} + \delta_{\rho} + \delta_{\varepsilon} \right)e^{I}, 
    \label{eq:difeoseasnueva}
  \\
  \delta_{\xi}\omega^{IJ}
  & ={\mathcal L}_{\xi} \omega^{IJ} 
  & = \left(\delta_{\tau} + \delta_{\rho} + \delta_{\varepsilon} \right)\omega^{IJ},
    \label{eq:difeoswasnueva} 
\end{eqnarray}
where $\xi:=\xi^{I}\partial_{I}$ is the infinitesimal generator of the diffeomorphism and the (field-dependent) gauge parameters are $\tau^{IJ}:= -\xi \intprod \omega^{IJ}$, $\rho^I := \xi \intprod e^I$ and $\varepsilon^{I} := \xi \intprod e^I$. Note that in three dimensions diffeomorphisms become a linear combination of local Lorentz transformations, local translations and the trivial symmetry \eref{eq:varenuevatrivial}-\eref{eq:varwnuevatrivial}.

We have seen in this section that the fundamental set of gauge symmetries of the Lovelock action \eref{eq:lovelockaction} for odd-dimensional manifolds $\mathcal{M}^{n}$ and the coefficients satisfy \eref{eq:sequencesolution} is formed by the gauge transformations $\delta_{\tau}$, $\delta_{\rho}$ and $\delta_{\varepsilon}$. Nevertheless, one can construct different complete sets of gauge symmetries which consider other symmetries as the fundamental ones. A particularly important aspect of these sets is their commutator algebra. So, for the sake of completeness, we report in what follows the commutator algebra of other two complete sets, equivalent to \eref{eq:LorentzLorentz0}-\eref{eq:transNewL0}, which are formed by the gauge symmetries obtained so far in this paper.

First set. This complete set of gauge symmetries is composed of local Lorentz transformations ($\delta_{\tau}$), local translations ($\delta_{\rho}$), and improved diffeomorphisms ($\delta_{\chi}$). The commutator algebra is
\begin{eqnarray}
  \fl \left[\delta_{\tau_{1}} , \delta_{\tau_{2}}\right] 
  & =\delta_{\tau_{3}}, 
  & \left(
    \tau_{3}^{IJ} := 2 {\tau_{1}}^{[ I \vert K}{\tau_{2}}^{\vert J]}{}_{K}
    \right),
    \label{eq:LorentzLorentz1}
  \\
  \fl \left[\delta_{\rho_{1}},\delta_{\rho_{2}}\right]
  & =\delta_{\tau}, 
  & \left(
    \tau^{IJ} := 2 \Lambda {\rho_{1}}^{[I}{\rho_{2}}^{J]}
    \right),
    \label{eq:transtrans1}
  \\
  \fl \left[\delta_{\tau},\delta_{\rho}\right]
  & =\delta_{\rho_{1}}, 
  & \left(
    {\rho_{1}}^{I} := -\tau^{I}{}_{J} \rho^{J}
    \right),
    \label{eq:Lorentztrans1}
  \\
  \fl \left[\delta_{\chi_{1}},\delta_{\chi_{2}}\right]
  & = \delta_{\tau} + \delta_{\chi_{3}},
  &  \left(
    \tau^{IJ} := \chi_{2} \intprod \left( \chi_{1}\intprod R^{IJ} \right),\quad
    \chi_{3}^{I} := \chi_{1} \intprod \left(\chi_{2}\intprod De^{I}\right)
    \right),
    \label{eq:DMDM}
  \\
  \fl \left[\delta_{\tau} , \delta_{\chi}\right]
  & =\delta_{\chi_{1}}, 
  & \left( 
    {\chi_{1}}^{I} := -\tau^{I}{}_{J}\chi^{J}
    \right),
    \label{eq:LorentzDM}
  \\
  \fl \left[\delta_{\rho},\delta_{\chi}\right]
  & =\delta_{\tau} + \delta_{\rho_{1}} +\delta_{\chi_{1}}, 
  & \left( 
    \tau^{IJ}:=2 \Lambda \rho^{[I} \chi^{J]},\right.
  \nonumber\\
  \fl
  & 
  & \left. \quad {\rho_{1}}^{I} := \chi \intprod D\rho^{I}, \quad 
    {\chi_{1}}^{I}:=-\chi \intprod D\rho^{I} \right).
    \label{eq:transDM}
\end{eqnarray}

Second set. This complete set of gauge symmetries is composed of local Lorentz transformations ($\delta_{\tau}$), local translations ($\delta_{\rho}$), and diffeomorphisms ($\delta_{\xi}$). The commutator algebra is
\begin{eqnarray}
  \fl \left[\delta_{\tau_{1}},\delta_{\tau_{2}}\right] 
  & =\delta_{\tau_{3}}, 
  & \left(
    {\tau_{3}}^{IJ} := 2{\tau_{1}}^{[ I \vert K}{\tau_{2}}^{\vert J]}{}_{K}
    \right),
    \label{eq:LorentzLorentz2}
  \\
  \fl \left[\delta_{\rho_{1}},\delta_{\rho_{2}}\right]
  &=\delta_{\tau}, 
  &  \left(
    \tau^{IJ} := 2 \Lambda {\rho_{1}}^{[I}{\rho_{2}}^{J]}
    \right),
    \label{eq:transtrans2}
  \\
  \fl \left[\delta_{\tau},\delta_{\rho}\right]
  & = \delta_{\rho_{1}}, 
  & \left(
    {\rho_{1}}^{I} := -\tau^{I}{}_{J}\rho^{J}
    \right),
    \label{eq:Lorentztrans2}
  \\
  \fl \left[\delta_{\xi_{1}},\delta_{\xi_{2}}\right]
  & = \delta_{\xi_{3}},
  & \left(
    {\xi_{3}}^{I}:=\xi_{1} \intprod \left(\xi_{2}\intprod de^{I}\right)
    \right),
    \label{eq:difdif}
  \\
  \fl \left[\delta_{\tau},\delta_{\xi}\right]
  & = \delta_{\tau} + \delta_{\xi_{1}}, 
  & \left(
    \tau^{IJ}:=\mathcal{L}_{\xi}\tau^{IJ}, \quad 
    {\xi_{1}}^{I} :=-\tau^{I}{}_{J}\xi^{J}
    \right),
    \label{eq:Lorentzdif}
  \\
  \fl \left[\delta_{\rho},\delta_{\xi}\right]
  & =\delta_{\rho_{1}} +\delta_{\xi_{1}}, 
  & \left( 
    {\rho_{1}}^{I} := \mathcal{L}_{\xi} \rho^{I},\quad 
    {\xi_{1}}^{I} := -\xi \intprod D\rho^{I} \right).
    \label{eq:transdif}
\end{eqnarray}

\section{A new gauge symmetry for the highest or last term of the Lovelock action in odd-dimensional manifolds $\mathcal{M}^{n}$}\label{sec:newsymmetry}
In this section we report the existence of a new symmetry that emerges from the application of the converse of the Noether's second theorem to the highest or last term of the Lovelock action \eref{eq:lovelockaction} for odd $n$, which is 
\begin{eqnarray}
  \label{eq:LovelockvanishingDeuler1}
  \int L_{n}^{(n-1)/2} = \kappa\int \epsilon_{I_{1}I_{2}\cdots I_{n-2}I_{n-1}I_{n}}
  R^{I_{1}I_{2}} \wedge \cdots \wedge R^{I_{n-2}I_{n-1}} \wedge e^{I_{n}}, 
\end{eqnarray}
with variational derivatives
\begin{eqnarray}
  \fl \mathcal{E}_{I}^{(n-1)/2} =
  & \kappa \epsilon_{I I_{2} I_{3} \cdots I_{n-1} I_{n}} R^{I_{2}I_{3}} \wedge \cdots \wedge R^{I_{n-1}I_{n}},
    \label{eq:eulere(n-1)/2}
  \\
  \fl \mathcal{E}_{IJ}^{(n-1)/2} =
  & \kappa \frac{(n-1)}{2} \epsilon_{I J I_{3} I_{4}\cdots I_{n-2} I_{n-1} I_{n}} R^{I_{3}I_{4}} \wedge \cdots \wedge R^{I_{n-2}I_{n-1}} \wedge De^{I_{n}}.
    \label{eq:eulerw(n-1)/2}
\end{eqnarray}
We have already seen in the subsection \ref{subsec:Diffeomorphisms} that the variational derivatives \eref{eq:eulere(n-1)/2} and \eref{eq:eulerw(n-1)/2}  are related by the Noether identity \eref{eq:Noetheridnew(n-1)/2}. Now we look for the gauge symmetry that emerges from this identity. Multiplying \eref{eq:Noetheridnew(n-1)/2} by the arbitrary local parameter $\varepsilon^{I}$ and denoting by $\varepsilon = \varepsilon^{I}\partial_{I}$ we get the off-shell identity
\begin{equation}
  \label{eq:Noetheridequivdiffeos(n-1)/2}
   \mathcal{E}^{(n-1)/2}_{I}
  \wedge \underbrace{\left(\varepsilon \intprod De^{I}\right)}_{\delta_{\varepsilon} e^{I}} +
  \mathcal{E}^{(n-1)/2}_{IJ} \wedge 
  \underbrace{ \left(\varepsilon \intprod R^{IJ} \right) }_{\delta_{\varepsilon}
    \omega^{IJ}} = 0.
\end{equation}
According to the converse of Noether's second theorem, the gauge symmetry involved in \eref{eq:Noetheridequivdiffeos(n-1)/2} is given by the terms accompanying the variational derivatives $\mathcal{E}^{(n-1)/2}_{I}$ and $\mathcal{E}^{(n-1)/2}_{IJ}$, which are
\begin{eqnarray}
  \delta_{\varepsilon} e^{I} 
  & = \varepsilon \intprod De^{I},
    \label{eq:varenueva}\\
  \delta_{\varepsilon} \omega^{IJ} 
  & = \varepsilon \intprod R^{IJ}.
  \label{eq:varwnueva}
\end{eqnarray}
This gauge symmetry had not been reported in the literature as far as we know, and it is remarkable that it is present for the action \eref{eq:LovelockvanishingDeuler1} in the odd-dimensional case only.

In fact, the action \eref{eq:LovelockvanishingDeuler1} is quasi-invariant under the new gauge symmetry \eref{eq:varenueva}-\eref{eq:varwnueva} because
\begin{eqnarray}
  \label{eq:variationNewLnot0}
  \delta S_{n}  
  = & -\kappa \int d \left( \frac12 (n-1)a_{(n-1)/2} \epsilon_{I J I_{3}I_{4}\cdots I_{n-2}I_{n-1} I_{n}}  
      \right. 
      \nonumber\\  
    &  \biggl. \times R^{I_{3}I_{4}} \wedge \cdots \wedge R^{I_{n-2}I_{n-1}} \wedge e^{I_{n}} \wedge \left(\varepsilon \intprod R^{IJ}\right) \biggr).
\end{eqnarray}

Note that in three dimensions this symmetry becomes trivial because the transformation \eref{eq:varenueva}-\eref{eq:varwnueva} can be rewritten in terms of the variational derivatives \eref{eq:eulere(n-1)/2}-\eref{eq:eulerw(n-1)/2} for $n=3$ as follows
\begin{eqnarray}
  \delta_{\varepsilon} e^{I} 
  & = \frac{\sigma}{2 \kappa a_{1} }\epsilon^{IJK} \left( \varepsilon \intprod \mathcal{E}^1_{JK}\right),
  \label{eq:varenuevatrivial0}\\
  \delta_{\varepsilon} \omega^{IJ} 
  & = \frac{ \sigma }{2 \kappa a_{1} }\epsilon^{IJK} \left( \varepsilon \intprod \mathcal{E}^1_{K}\right).
  \label{eq:varwnuevatrivial0}
\end{eqnarray}
Nevertheless, we see from \eref{eq:eulere(n-1)/2}-\eref{eq:eulerw(n-1)/2} that in dimensions higher than three the new symmetry \eref{eq:varenueva}-\eref{eq:varwnueva} is \textit{not} trivial.

Therefore, the complete set of gauge symmetries of the action \eref{eq:LovelockvanishingDeuler1} is composed of local Lorentz transformations, local translations with $\Lambda=0$, and also the new gauge symmetry \eref{eq:varenueva}-\eref{eq:varwnueva}. Now we obtain the algebra of the gauge symmetries of \eref{eq:LovelockvanishingDeuler1} by computing the commutators among their respective variations, acting on both the frame and the connection. Denoting a local Lorentz transformation by $\delta_{\tau}$, a local translation (with $\Lambda=0$) by $\delta_{\rho}$ and the symmetry \eref{eq:varenueva}-\eref{eq:varwnueva} by $\delta_{\varepsilon}$, the algebra reads
\begin{eqnarray}
  \fl \left[\delta_{\tau_{1}},\delta_{\tau_{2}}\right]
  & =\delta_{\tau_{3}}, 
  & \left(
    {\tau_{3}}^{IJ} := 2{\tau_{1}}^{[ I \vert K}{\tau_{2}}^{\vert J]}{}_{K}
    \right),
    \label{eq:LorentzLorentz}
  \\
  \fl \left[\delta_{\rho_{1}},\delta_{\rho_{2}}\right]
  & =0,
  & 
    \label{eq:transtrans}
  \\
  \fl \left[\delta_{\tau},\delta_{\rho}\right]
  & =\delta_{\rho_{1}},
  &\left(
    {\rho_{1}}^{I} := - {\tau^{I}}_{J}\rho^{J}
    \right),
    \label{eq:Lorentztrans}
  \\
  \fl \left[\delta_{\varepsilon_{1}},\delta_{\varepsilon_{2}}\right]
  & = \delta_{\tau} + \delta_{\rho} + \delta_{\varepsilon_{3}},
  & \left(
    \tau^{IJ} := \varepsilon_{2} \intprod \left( \varepsilon_{1}\intprod R^{IJ} \right),
    \quad \rho^{I}:= \varepsilon_{1} \intprod \left(\varepsilon_{2}\intprod De^{I}\right)\right.,
    \nonumber\\
  \fl        
  &  
  & \quad \left. 
    {\varepsilon_{3}}^{I} := \varepsilon_{1} \intprod \left(\varepsilon_{2}\intprod De^{I}\right)
    + \varepsilon_{2}\intprod D {\varepsilon_{1}}^{I} - \varepsilon_{1} \intprod D {\varepsilon_{2}}^{I}
    \right),
    \label{eq:newnew}
  \\
  \fl \left[\delta_{\tau},\delta_{\varepsilon}\right]
  & =\delta_{\varepsilon_{1}}, 
  &\left( 
    {\varepsilon_{1}}^{I} := -\tau^{I}{}_{J}\varepsilon^{J}
    \right),
    \label{eq:Lorentznew}
  \\
  \fl \left[\delta_{\rho},\delta_{\varepsilon}\right]
  &=\delta_{\varepsilon_{1}},
  &\left( 
    {\varepsilon_{1}}^{I} := -\varepsilon \intprod D\rho^{I}
    \right).
    \label{eq:transnew}
\end{eqnarray}
First of all, we observe that the algebra of the gauge symmetries of can be obtained from \eref{eq:LorentzLorentz0}-\eref{eq:transNewL0} by simply setting $\Lambda=0$ (although they correspond to completely different choices of the coefficients $a_{p}$ in Lovelock action). The equations \eref{eq:LorentzLorentz}-\eref{eq:Lorentztrans} reflect the fact that the commutators of local Lorentz transformations and local translations together form the algebra of the Poincar\'e or Euclidean groups. On the other hand, \eref{eq:newnew} states that the commutator of two new symmetries \eref{eq:varenueva}-\eref{eq:varwnueva} is a linear combination of a local Lorentz transformation, a local translation and a transformation of the same type as \eref{eq:varenueva}-\eref{eq:varwnueva}, with different field-dependent gauge parameters. Finally, the relationship \eref{eq:Lorentznew} (\eref{eq:transnew}, respectively) reveals that the commutator of a local Lorentz transformation (local translation) with the new gauge symmetry is again a transformation of the type of \eref{eq:varenueva}-\eref{eq:varwnueva}, with a rotated (translated) gauge parameter.

As we have seen in the subsection \eref{subsec:Diffeomorphisms}, the action \eref{eq:LovelockvanishingDeuler1} is also invariant under diffeomorphisms. However, we have shown that the set of gauge symmetries of \eref{eq:LovelockvanishingDeuler1} is complete (because the algebra closes). Therefore, diffeomorphisms become a derived symmetry. In fact, if $\xi$ is any vector field, then an infinitesimal diffeomorphism can be expressed in terms of the gauge symmetries that form the complete set as
\begin{eqnarray}
  \delta_{\xi}e^{I}
  & ={\mathcal L}_{\xi }e^{I} 
  & = \left(\delta_{\tau} + \delta_{\rho} + \delta_{\varepsilon} \right)e^{I}, 
    \label{eq:difeoseasnueva}
  \\
  \delta_{\xi}\omega^{IJ}
  & ={\mathcal L}_{\xi} \omega^{IJ} 
  & = \left(\delta_{\tau} + \delta_{\rho} + \delta_{\varepsilon} \right)\omega^{IJ},
    \label{eq:difeoswasnueva}
\end{eqnarray}
with field-dependent gauge parameters $\tau^{IJ}:= -\xi \intprod \omega^{IJ}$, $\rho^I := \xi \intprod e^I$ and $\varepsilon^{I} := \xi \intprod e^I$.

\section{Conclusion}\label{sec:Conclusion}
In this paper we have applied the converse of Noether's second theorem to the first-order Lovelock action for gravity, obtaining in first instance its well-known gauge symmetries under local Lorentz transformations and diffeomorphisms. After that, we have shown that for odd dimensions there is a generalization of the so-called three-dimensional `local translations', that turns out to be non-equivalent to the diffeomorphisms in dimensions higher than three. Thus, the gauge invariance of Lovelock theory can be described by the complete set of gauge symmetries composed of local Lorentz transformations, diffeomorphisms and when the coefficients $a_{p}$ satisfy the relation \eref{eq:sequencesolution}  or consists of the single term \eref{eq:LovelockvanishingDeuler1}, also local translations. 

After that, also in the odd-dimensional case, we have obtained a new gauge symmetry of the Lovelock action when its coefficients satisfy the relation \eref{eq:sequencesolution}. This symmetry results to be trivial only for $n=3$. Therefore, the complete set of gauge symmetries of the Lovelock action in odd dimensions ($n>3$) with this particular choice of coefficients can be considered as the (anti-)de Sitter transformations together with the new symmetry \eref{eq:varenuevaL}-\eref{eq:varwnuevaL}. We have also computed the commutator algebra of this complete set of gauge symmetries and, for the sake of completeness, the commutator algebras of other two equivalent complete sets of gauge symmetries, obtaining that in all these cases the algebras close with structure functions.

On the other hand, for odd-dimensional manifolds $\mathcal{M}^{n}$ and when the Lovelock action is solely given by the single term \eref{eq:LovelockvanishingDeuler1}, we have obtained a new gauge \eref{eq:varenueva}-\eref{eq:varwnueva}, which is analogous to \eref{eq:varenuevaL}-\eref{eq:varwnuevaL} . The complete set of gauge symmetries of this action is composed of Poincar\'e ($\sigma=-1$) or Euclidean ($\sigma=1$) transformations and the new symmetry \eref{eq:varenueva}-\eref{eq:varwnueva}. We have also reported the commutator algebra of the gauge symmetries of this last action, which turns out to be closed with structure functions.

Furthermore, the new gauge symmetries \eref{eq:varenuevaL}-\eref{eq:varwnuevaL} and \eref{eq:varenueva}-\eref{eq:varwnueva} could also be used to connect mathematically different solutions, but in fact related by a gauge transformation and therefore physically equivalent. Additionally, to find the generators of these symmetries could be useful for the quantization programme of Lovelock gravity.

We conclude by pointing out that it would be interesting to study other models of gravity from the perspective of this work because, as we have seen here, the application of the converse of Noether's second theorem can uncover new gauge symmetries of such models or at least lead to a reformulation of the existing ones finding new complete sets of gauge symmetries. In particular, to look for the existence of the analog of local translations in other theories different from general relativity might be particularly interesting.

\ack

We thank Mariano Celada, Diego Gonz\'alez, and Ricardo Troncoso for their valuable comments.  Bogar D\'{i}az is supported with a VIEP-BUAP postdoctoral fellowship.

\appendix
\section{Equivalent complete sets of gauge symmetries and their
  corresponding commutator algebras for odd-dimensional manifolds
  $\mathcal{M}^{n}$ with generic gauge parameters}\label{app:completealgebra}

By considering field-independent gauge parameters--because that is the `natural' choice--in section \ref{subsec:NewsymmetryLnot0} we computed the commutator algebra of three equivalent complete sets of gauge symmetries of the Lovelock action \eref{eq:lovelockaction} with coefficients satisfying \eref{eq:sequencesolution}. However, the parameters that multiply the Noether identities considered in the converse of Noether's second theorem might also depend on the fields (and its derivatives). Therefore, for the sake of completeness, in this appendix we compute the commutator algebras in the most general case, which allows field-dependent gauge parameters. It is important to remark that regardless of the choice of the gauge parameters, the transformation law of the frame and the connection is always the same. 

First set. This complete set of gauge symmetries is composed of local Lorentz transformations ($\delta_{\tau}$), local translations ($\delta_{\rho}$), and improved diffeomorphisms ($\delta_{\chi}$) with infinitesimal generator $\chi := \chi^{I} \partial_{I}$.  The commutator algebra is
\begin{eqnarray}
  \fl\left[\delta_{\tau_{1}} , \delta_{\tau_{2}}\right] 
  & = \delta_{\tau_{3}},
  & \left(
    {\tau_{3}}^{IJ}:= \delta_{\tau_{1}}{\tau_{2}}^{IJ} - \delta_{\tau_{2}}{\tau_{1}}^{IJ} + 2{\tau_{1}}^{[ I \vert K}{{\tau_{2}}^{ \vert J ]}}_{K}
    \right),
    \label{eq:LorentzLorentzg1}
  \\
  \fl\left[\delta_{\rho_{1}} , \delta_{\rho_{2}}\right] 
  & = \delta_{\tau} + \delta_{\rho_{3}},
  & \left(
    {\tau}^{IJ} := 2 \Lambda {\rho_{1}}^{[I}{\rho_{2}}^{J]},
    {\rho_{3}}^{I} := \delta_{\rho_{1}}{\rho_{2}}^{I}-\delta_{\rho_{2}}{\rho_{1}}^{I}
    \right),
    \label{eq:trtrg1}
  \\
  \fl\left[\delta_{\tau} , \delta_{\rho}\right] 
  & = \delta_{\tau_{1}} + \delta_{\rho_{1}},
  & \left(
    {\tau_{1}}^{IJ} := -\delta_{\rho}\tau^{IJ}, \quad
    {\rho_{1}}^{I} := \delta_{\tau}\rho^{I} - {\tau^{I}}_{J}\rho^{J}
    \right),
    \label{eq:Lorentztr1}
  \\
    \fl\left[\delta_{\chi_{1}} , \delta_{\chi_{2}}\right]
  & = \delta_{\tau} + \delta_{\chi_{3}},
  & \left(
    \tau^{IJ} := \chi_{2}\intprod \left(\chi_{1}\intprod R^{IJ}\right),
    \right.\nonumber
  \\
  \fl
  &
  & \quad \left.
    {\chi_{3}}^{I} := \left(\delta_{\chi_{1}}\chi_{2}-\delta_{\chi_{2}}\chi_{1}-\left[\chi_{1},\chi_{2}\right]\right) \intprod e^{I}
    \right),
    \label{eq:DMDMg1}
  \\
  \fl\left[ \delta_{\tau} , \delta_{\chi}\right]
  & = \delta_{\tau_{1}} + \delta_{\chi_{1}},
  & \left(
    {\tau_{1}}^{IJ} := -\delta_{\chi}\tau^{IJ} , \quad
    {\chi_{1}}^{I} :=   \left( \delta_{\tau}\chi \right)\intprod e^{I}
    \right),
    \label{eq:LorentzDMg1}
  \\
  \fl \left[\delta_{\rho} , \delta_{\chi}\right]
  & = \delta_{\rho_{1}} + \delta_{\chi_{1}},
  & \left(
    {\rho_{1}}^{I} := \mathcal{L}_{\chi}\rho^{I} + \left(\chi \intprod {\omega^{I}}_{J}\right)\rho^{J}-\delta_{\chi}\rho^{J},
    \right.\nonumber
  \\
  \fl
  &
  & \left. 
    \quad {\chi_{1}}^{I} := \left( \delta_{\rho}\chi \right)\intprod e^{I}
    \right).
    \label{eq:trDMg1}
\end{eqnarray}
Note that $\delta_{\chi_{1}}\chi_{2} = \left(\delta_{\chi_{1}}{\chi_{2}}^{I}\right) \partial_{I} + {\chi_{2}}^{I} \left(\delta_{\chi_{1}}\partial_{I}\right)$, and so on.

Second set. This complete set of gauge symmetries is composed of local Lorentz transformations ($\delta_{\tau}$), local translations ($\delta_{\rho}$), and diffeomorphisms ($\delta_{\xi}$) with infinitesimal generator $\xi:=\xi^{I} \partial_{I}$. The commutator algebra is
\begin{eqnarray}
  \fl\left[\delta_{\tau_{1}} , \delta_{\tau_{2}}\right] 
  & = \delta_{\tau_{3}}, \quad
  & \left(
    {\tau_{3}}^{IJ}:= \delta_{\tau_{1}}{\tau_{2}}^{IJ} - \delta_{\tau_{2}}{\tau_{1}}^{IJ} + 2{\tau_{1}}^{[ I \vert K}{{\tau_{2}}^{ \vert J ]}}_{K}
    \right),
    \label{eq:LorentzLorentzg2}
  \\
  \fl\left[\delta_{\rho_{1}} , \delta_{\rho_{2}}\right] 
  & = \delta_{\tau} + \delta_{\rho_{3}}, \quad
  & \left(
    {\tau}^{IJ} := 2 \Lambda {\rho_{1}}^{[I}{\rho_{2}}^{J]}, \quad
    {\rho_{3}}^{I} := \delta_{\rho_{1}}{\rho_{2}}^{I}-\delta_{\rho_{2}}{\rho_{1}}^{I}
    \right),
    \label{eq:ltltg2}
  \\
  \fl \left[\delta_{\tau} , \delta_{\rho}\right] 
  & = \delta_{\tau_{1}} + \delta_{\rho_{1}}, \quad
  & \left(
    {\tau_{1}}^{IJ} := -\delta_{\rho}\tau^{IJ}, \quad
    {\rho_{1}}^{I} := \delta_{\tau}\rho^{I} - {\tau^{I}}_{J}\rho^{J}
    \right),
    \label{eq:Lorentzltg2}
  \\
  \fl\left[\delta_{\xi_{1}} , \delta_{\xi_{2}}\right]
  & = \delta_{\xi_{3}},
  & \left(
    {\xi_{3}}^{I} := \left(\delta_{\xi_{1}}\xi_{2}-\delta_{\xi_{2}}\xi_{1}-\left[\xi_{1},\xi_{2}\right]\right) \intprod e^{I}
    \right),
    \label{eq:difdifg2}
  \\
  \fl \left[\delta_{\tau} , \delta_{\xi} \right] 
  & = \delta_{\tau_{1}} + \delta_{\xi_{1}}, \quad
  & \left(
    {\tau_{1}}^{IJ} := \mathcal{L}_{\xi}\tau^{IJ} - \delta_{\xi}\tau^{IJ},\quad
    {\xi_{1}}^{I} := \left(\delta_{\tau} \xi \right) \intprod e^{I}
    \right),
    \label{eq:Lorentzdifg2}
  \\
  \fl\left[\delta_{\rho} , \delta_{\xi}\right]
  & = \delta_{\rho_{1}} + \delta_{\xi_{1}} ,\quad
  &\left(
    {\rho_{1}}^{I}:=\mathcal{L}_{\xi}\rho^{I} - \delta_{\xi}\rho^{I},\quad
    {\xi_{1}}^{I}:=\left(\delta_{\rho}\xi\right) \intprod e^{I}
    \right).
    \label{eq:ltdifg2}
\end{eqnarray}
Observe that $\delta_{\xi_{1}}\xi_{2} = \left(\delta_{\xi_{1}}{\xi_{2}}^{I}\right) \partial_{I} + {\xi_{2}}^{I} \left(\delta_{\xi_{1}}\partial_{I}\right)$, and so on. 

Third set. This complete set of gauge symmetries is composed of local Lorentz transformations ($\delta_{\tau}$), local translations ($\delta_{\rho}$), and the new gauge symmetry ($\delta_{\varepsilon}$) with  $\varepsilon := \varepsilon^{I} \partial_{I}$. The commutator algebra is
\begin{eqnarray}
  \fl \left[\delta_{\tau_{1}} , \delta_{\tau_{2}}\right] 
  & = \delta_{\tau_{3}}, \quad
  & \left(
    {\tau_{3}}^{IJ}:= \delta_{\tau_{1}}{\tau_{2}}^{IJ} - \delta_{\tau_{2}}{\tau_{1}}^{IJ} + 2{\tau_{1}}^{[ I \vert K}{{\tau_{2}}^{ \vert J ]}}_{K}
    \right),
  \\
  \fl \left[\delta_{\rho_{1}} , \delta_{\rho_{2}}\right] 
  & = \delta_{\tau} + \delta_{\rho_{3}}, \quad
  & \left(
    {\tau}^{IJ} := 2 \Lambda {\rho_{1}}^{[I}{\rho_{2}}^{J]}, \quad
    {\rho_{3}^{I}} := \delta_{\rho_{1}}{\rho_{2}}^{I}-\delta_{\rho_{2}}{\rho_{1}}^{I}
    \right),
  \\
  \fl \left[\delta_{\tau} , \delta_{\rho}\right] 
  & = \delta_{\tau_{1}} + \delta_{\rho_{1}}, \quad
  & \left(
    {\tau_{1}}^{IJ} := -\delta_{\rho}\tau^{IJ}, \quad
    {\rho_{1}}^{I} := \delta_{\tau}\rho^{I} - {\tau^{I}}_{J}\rho^{J}
    \right),
  \\
  \fl \left[\delta_{\varepsilon_{1}} , \delta_{\varepsilon_{2}}\right]
  & = \delta_{\tau} + \delta_{\rho} + \delta_{\varepsilon_{3}},
  & \left(
    \tau^{IJ}:=\varepsilon_{1}\intprod \left( \varepsilon_{2} \intprod R^{IJ} \right) + 2 \Lambda {\varepsilon_{1}}^{[I}{\varepsilon_{2}}^{J]},
    \right.\nonumber
  \\
  \fl
  &
  &  \quad \rho^{I} := \varepsilon_{1} \intprod \left( \varepsilon_{2} \intprod De^{I}\right),\nonumber
  \\
  \fl
  &
  & \quad \left. 
    {\varepsilon_{3}}^{I} := \left( \delta_{\varepsilon_{1}} \varepsilon_{2}-\delta_{\varepsilon_{2}} \varepsilon_{1}
    -\left[\varepsilon_{1},\varepsilon_{2}\right] \right) \intprod e^{I}
    \right),
  \\
  \fl \left[ \delta_{\tau} , \delta_{\varepsilon} \right]
  & = \delta_{\tau_{1}} + \delta_{\varepsilon_{1}}, 
  &\left(
    {\tau_{1}}^{IJ} := -\delta_{\varepsilon} \tau^{IJ} ,\quad
    {\varepsilon_{1}}^{I} := \left( \delta_{\tau} \varepsilon \right) \intprod e^{I}
    \right),
  \\
  \fl \left[\delta_{\rho} , \delta_{\varepsilon}\right]
  & = \delta_{\rho_{1}} + \delta_{\varepsilon_{1}} , 
  & \left(
    {\rho_{1}}^{I} := - \delta_{\varepsilon} \rho^{I},\quad
    {\varepsilon_{1}}^{I} := \left( \delta_{\rho}\varepsilon\right) \intprod e^{I}
    \right).
\end{eqnarray}
Notice that $\delta_{\varepsilon_{1}}\varepsilon_{2} = \left(\delta_{\varepsilon_{1}}{\varepsilon_{2}}^{I}\right) \partial_{I} + {\varepsilon_{2}}^{I} \left(\delta_{\varepsilon_{1}}\partial_{I}\right)$, and so on.

Therefore, if the gauge parameters $\tau^{IJ}$, $\rho^I$, $\chi^I$, $\xi^I$, and $\varepsilon^I$ are taken  
field-independent, then these commutator algebras become those reported in the sections \ref{subsec:NewsymmetryLnot0} and \ref{sec:newsymmetry} of this paper. 

However, other choices where the gauge parameters are field-dependent are also allowed and frequently used in the literature. Let us illustrate one of them as an example. For instance, in \eref{eq:difdifg2} we need to compute $\delta_{\xi_{2}}{\xi_{1}}$. Therefore, using ${\xi_{1}}^I \partial_{I}$ we get
\begin{eqnarray}
\left ( \delta_{\xi_{2}}{\xi_{1}}^{I} \right ) \partial_I  = \delta_{\xi_{2}} \xi_{1}  + \left( \xi_{1} \intprod \delta_{\xi_{2}} e^{I}\right) \partial_{I}.
\end{eqnarray}
Thus, if we make the choice $\delta_{\xi_{2}} \xi_{1} = 0$ and take into account that $\delta_{\xi_{2}} e^{I} =\mathcal{L}_{\xi_2} e^I$, we get
\begin{eqnarray}
\delta_{\xi_{2}}{\xi_{1}}^{I} =  \xi_{1} \intprod \left( \mathcal{L}_{\xi_2} e^I \right),
\end{eqnarray}
which clearly shows that the gauge parameters $\xi^{I}$ are field-dependent. Therefore, with such a choice, \eref{eq:difdifg2} becomes
\begin{eqnarray}
 \left[\delta_{\xi_{1}} , \delta_{\xi_{2}}\right] = \delta_{\xi_{3}}, \quad  {\xi_{3}}^{I}  := \left[\xi_{2},\xi_{1}\right] \intprod e^{I}.
 \label{eq:difdifRH}
\end{eqnarray}

On the other hand, in \eref{eq:Lorentzdifg2} the terms $\delta_{\xi}\tau^{IJ}$ and $ \delta_{\tau} \xi$ must be computed. Using $\xi = \xi^{I}\partial_{I}$, we get
\begin{eqnarray}
  \left ( \delta_{\tau}{\xi}^{I} \right ) \partial_I  = \delta_{\tau} \xi_{1}  + \left( \xi \intprod \delta_{\tau} e^{I}\right) \partial_{I}.
\end{eqnarray}
Therefore, by making the choice $\delta_{\tau} \xi = 0$ and recalling that $\delta_{\tau} e^{I} =\tau^{I}{}_{J} e^{J}$, we obtain 
\begin{eqnarray}
\delta_{\tau}\xi^{I} =  \tau^{I}{}_{J}\xi^{J},
\end{eqnarray}
which means that the gauge parameters $\xi^{I}$ are field-dependent (and in fact they transform as the components of a Lorentz vector). Additionally, we can take $\tau^{IJ}$ as field-independent gauge parameters and, with all these choices, \eref{eq:Lorentzdifg2} becomes
\begin{eqnarray}
 \left[\delta_{\tau} , \delta_{\xi} \right] 
  & = \delta_{\tau_{1}},\quad
  & {\tau_{1}}^{IJ} := \mathcal{L}_{\xi}\tau^{IJ}.
    \label{eq:LorentzdifRH}
\end{eqnarray}

Finally, in \eref{eq:LorentzLorentzg2} we can choose the gauge parameters ${\tau_{1}}^{IJ}$ and ${\tau_{2}}^{IJ}$ as field-independent because this is a common choice in the literature. This choice leads to $\delta_{\tau_{1}}{\tau_{2}}^{IJ} = \delta_{\tau_{2}}{\tau_{1}}^{IJ} = 0$, and in this case the commutator \eref{eq:LorentzLorentzg2} reduces to
\begin{eqnarray}
    \left[\delta_{\tau_{1}} , \delta_{\tau_{2}}\right] 
  & = \delta_{\tau_{3}}, \quad
  & {\tau_{3}}^{IJ}:= 2{\tau_{1}}^{[ I \vert K}{{\tau_{2}}^{ \vert J ]}}_{K}.
    \label{eq:LorentzLorentzRH}
\end{eqnarray}

Summarizing, if we take the gauge parameters $\tau^{IJ}$ and the vector field $\xi = \xi^{I}\partial_{I}$ as field-independent (which amounts to take the parameters $\xi^{I}$ field-dependent), the commutator algebra of local Lorentz transformations and diffeomorphisms becomes \eref{eq:difdifRH}, \eref{eq:LorentzdifRH} and \eref{eq:LorentzLorentzRH}.

\section*{References}
\bibliography{Bibliography}

\end{document}